\documentclass[sigconf]{acmart}

\AtBeginDocument{%
  \providecommand\BibTeX{{%
    \normalfont B\kern-0.5em{\scshape i\kern-0.25em b}\kern-0.8em\TeX}}}

\copyrightyear{2022} 
\acmYear{2022} 
\setcopyright{acmcopyright}\acmConference[CIKM '22]{Proceedings of the 31st ACM International Conference on Information and Knowledge Management}{October 17--21, 2022}{Atlanta, GA, USA}
\acmBooktitle{Proceedings of the 31st ACM International Conference on Information and Knowledge Management (CIKM '22), October 17--21, 2022, Atlanta, GA, USA}
\acmPrice{15.00}
\acmDOI{10.1145/3511808.3557415}
\acmISBN{978-1-4503-9236-5/22/10}

\usepackage{tcolorbox}

\usepackage{multirow}

\usepackage[normalem]{ulem}
\useunder{\uline}{\ul}{}

\usepackage{pgfplots}
\usepgfplotslibrary{groupplots}
\pgfplotsset{compat=1.14}
\usepackage{tikzpeople}

\usepackage{graphicx}

\usepackage{sistyle}
\SIthousandsep{,}

\usepackage{makecell}

\usepackage{caption}
\usepackage{subcaption}

\usepackage{mathtools}

\usepackage{amsmath}

\usepackage{pifont}

\usepackage{cleveref}

\DeclareMathAlphabet{\mathbbb}{U}{bbold}{m}{n}

\crefformat{section}{\S#2#1#3} 
\crefformat{subsection}{\S#2#1#3}
\crefformat{subsubsection}{\S#2#1#3}

\definecolor{gred}{RGB}{219,68,55}
\definecolor{gblue}{RGB}{66,133,244}

\newcommand{\cmark}{\color{gblue}\ding{51}}
\newcommand{\xmark}{\color{gred}\ding{55}}

\newcommand{\MD}{{\small\textsf{PERIS}}}

\DeclarePairedDelimiter\ceil{\lceil}{\rceil}

\begin{document}

\title{Beyond Learning from Next Item: Sequential Recommendation via Personalized Interest Sustainability}


\author{Dongmin Hyun}
\affiliation{%
  \institution{Pohang University of Science and Technology}
  \city{Pohang}
  \country{Republic of Korea}
}
\email{dm.hyun@postech.ac.kr}

\author{Chanyoung Park}
\affiliation{%
  \institution{Korea Advanced Institute of Science and Technology}
  \city{Daejeon}
  \country{Republic of Korea}}
\email{cy.park@kaist.ac.kr}

\author{Junsu Cho}
\affiliation{%
  \institution{Pohang University of Science and Technology}
  \city{Pohang}
  \country{Republic of Korea}
}
\email{junsu7463@postech.ac.kr}

\author{Hwanjo Yu$^*$}
\affiliation{%
  \institution{Pohang University of Science and Technology}
  \city{Pohang}
  \country{Republic of Korea}
}
\email{hwanjoyu@postech.ac.kr}

\renewcommand{\shortauthors}{Dongmin Hyun, Chanyoung Park, Junsu Cho, and Hwanjo Yu}

\begin{abstract} 
Sequential recommender systems have shown effective suggestions by capturing users' interest drift. 
There have been two groups of existing sequential models: user- and item-centric models.
The user-centric models capture \textit{personalized} interest drift based on each user's sequential consumption history, but do not explicitly consider whether users’ interest in items sustains beyond the training time, i.e., interest \textit{sustainability}. On the other hand, the item-centric models consider whether users’ general interest sustains after the training time,
but it is not personalized. 
In this work, we propose a recommender system taking advantages of the models in both categories. Our proposed model captures \textit{personalized} interest \textit{sustainability}, indicating whether each user’s interest in items will sustain beyond the training time or not. 
We first formulate a task that requires to predict which items each user will consume in the recent period of the training time based on users' consumption history.
We then propose simple yet effective schemes to augment users' sparse consumption history. 
Extensive experiments show that the proposed model outperforms 10 baseline models on 11 real-world datasets. The codes are available at: \href{https://github.com/dmhyun/PERIS}{\color{magenta}{https://github.com/dmhyun/PERIS}}.
\end{abstract}




\maketitle

\begin{figure}[t]
	\centering
	\includegraphics[width=.999\linewidth]{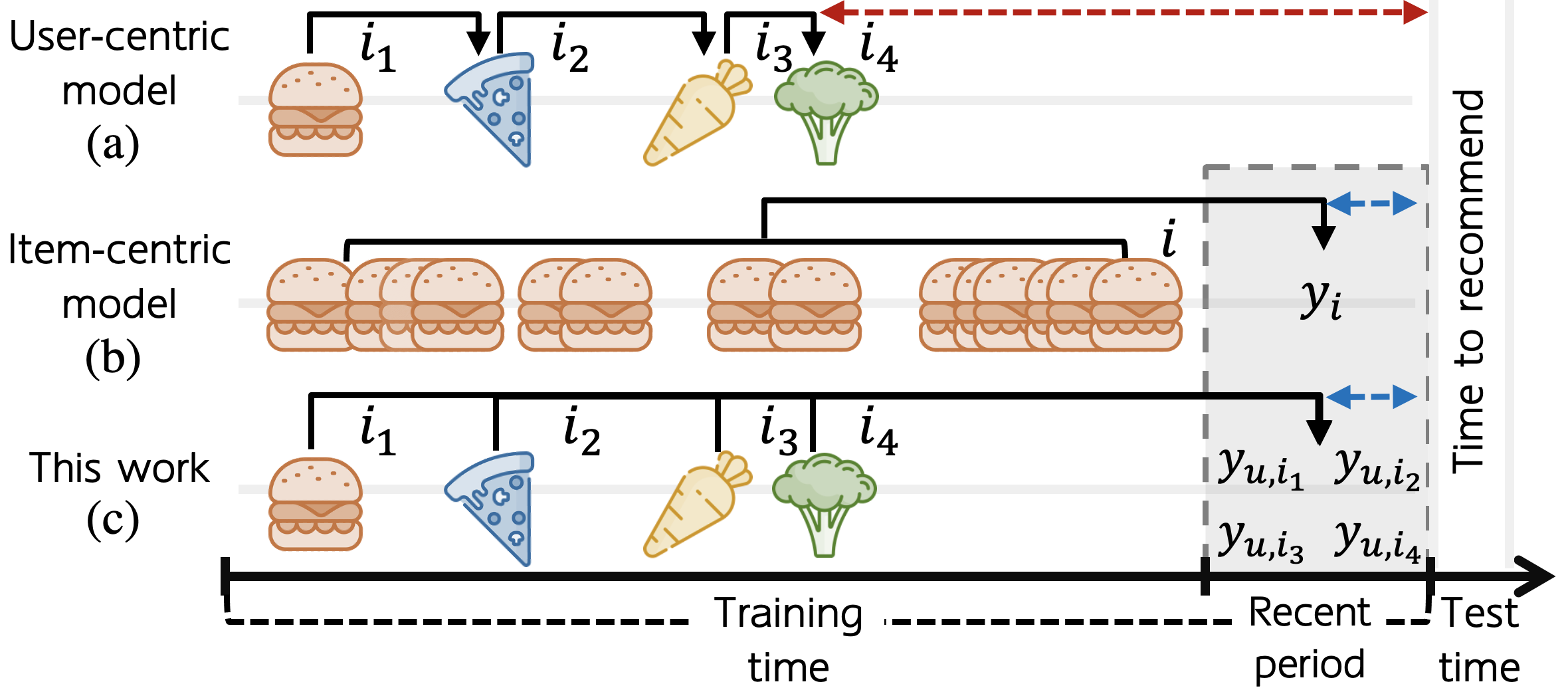}
	\caption{Comparison of models capturing interest drift. Solid arrow denotes prediction based on items.
	Labels $y_i$ and $y_{u,i}$: whether any user consumed item $i$ and whether a user $u$ consumed item $i$, in the recent period, respectively. Dashed arrow denotes the elapsed time from the last prediction.}
	\label{fig:intro}
\end{figure}

\section{Introduction}
Recommender system has been an indispensable technique to provide users with appealing items (e.g., products or services) from large catalogs of candidate items \cite{adomavicius2005toward}. Recent research has focused on sequential recommender system that captures users' interest drifting from the past to the recent to accurately suggest attractive items based on users' sequential history of consumption \cite{kang2018self,ma2019hierarchical,hyun2020interest,li2021lightweight}. 
There have been two groups of the sequential models capturing interest drift based on how they utilize the sequential history of consumption:  {\romannumeral 1}) user- and {\romannumeral 2}) item-centric models.  

The user-centric models capture the interest drift in items for \textit{each user} based on each user's chronological sequence of consumed items \cite{ma2019hierarchical,li2021lightweight} (Figure \ref{fig:intro}a). Thus, the user-centric model can track \textit{personalized} interest drift over time. 
However, the user-centric models do not explicitly consider whether users’ interest sustains beyond the training time. 
Concretely, user-centric models learn user representations based on the next item prediction (Figure \ref{fig:intro}a), and thus the user representation reflects the user's interest only up to the user's last consumption. For example, the representation of a user who consumed items up to \textit{2019} contains the user's interest only up to \textit{2019}, which can be inaccurate to perform  recommendation in \textit{2022} (i.e., after the training time) due to the prolonged absence of consumption, e.g., a long red line in Figure \ref{fig:intro}a. 

The item-centric models utilize all users' consumption history for each item to capture users' general interest in \textit{each item} \cite{wang2019modeling,hyun2020interest} (Figure \ref{fig:intro}b). In a recent work \cite{hyun2020interest}, the model captures whether users' general interest in each item will sustain beyond the training time, and the notion is called interest \textit{sustainability}. Specifically, it explicitly predicts whether each item will be consumed in a recent period of training data, i.e., the gray box in Figure \ref{fig:intro}b. 
Thus, users' interest learned by the item-centric model can align better to the users' interest in the test time, i.e., the future, than the user-centric models thanks to the shortened period of time between the time at which users' last interest is captured and the test time, i.e., a short blue line in Figure \ref{fig:intro}b.
However, the model only learns whether users' non-personalized interest sustains beyond the training time, and thus the item-centric model assigns the same interest sustainability for items even to users with different tastes, e.g., recommending generally-consumed coffee to a person who has caffeine allergy.

Motivated by these limitations, we propose a recommender system that takes the benefits of both user- and item-centric models while addressing  the downsides. 
Our method, \underline{Per}sonalized \underline{I}nterest \underline{S}ustainability-aware recommender system (\MD), learns each user's interest sustainability by predicting which items each user will consume in the recent period of the training time, i.e., the gray box in Figure \ref{fig:intro}c.
As a result, \MD{} can learn the personalized interest sustainability for items by considering \textit{each user}'s consumption in the recent period of training data, which cannot be learned by either the user-centric or item-centric models. 
However, it is non-trivial to predict items that each user is likely to consume in the recent period of the training time because most users have insufficient consumption history per item, e.g., users have 2.6 interactions per item on average in Yelp data.
To this end, we devise simple yet effective schemes to supplement users' sparse consumption history in both intrinsic and extrinsic manners.

The intrinsic scheme augments each user's consumption history for an item based on \textit{other items consumed by the user}. 
Its underlying idea is that a user's interest in an item (e.g., \textit{espresso}) is assumed to sustain if the user recently consumes a similar item (e.g., \textit{cappuccino}).
Hence, the intrinsic scheme is beneficial to supplement each user's consumption history for an item if the user consumes a variety of items. 
In addition, we devise the extrinsic scheme to supplement a target user's consumption history by referring \textit{other like-minded users}' consumption history.
The idea is that we can infer the interest of a target user (e.g., \textit{vegetarian}) in items (e.g., \textit{foods}) through the interest of like-minded users (e.g., \textit{other vegetarians}) in those items. 
Specifically, the extrinsic scheme trains the model to predict like-minded users' interest in the future to infer a target user's interest. 

Experiments show that \MD{} outperforms 10 baseline recommender systems such as general, user-centric, and item-centric models on 11 real-world datasets. 
In addition, we observe that \MD{} consistently enhances the recommendation accuracy over different elapsed times since users' last consumption compared to the baseline models, implying the personalized interest sustainability is beneficial to accurately infer users’ interest drift.
Moreover, we observe that \MD{} successfully captures the personalized interest sustainability, which is not fully captured by existing user- and item-centric sequential recommender systems. 

\section{Related Work}

\subsection{\textbf{General Recommender Systems}}
The general recommender systems learn each user's preference from a set of consumed items, i.e., no order information among consumed items. 
Bayesian personalized ranking (BPR) \cite{rendle2012bpr} formulates a pair-wise ranking loss to train recommender systems.
CML \cite{hsieh2017collaborative} adopts the metric learning to train a recommender system for satisfying the triangle inequality, which cannot be satisfied by the widely-used inner product operation.
TransCF \cite{park2018collaborative} extends CML by applying translation vectors to users and items. 
SML \cite{li2020symmetric} also enhances CML by incorporating an item-side training objective and trainable parameters for margins.
SimpleX \cite{mao2021simplex} is a model based on a contrastive learning and it outperforms recent recommender systems including the models based on the metric learning.
However, the general models do not utilize the order information in users' consumption history to track their interest drift.

\subsection{Sequential Recommender Systems} 
\subsubsection{\textbf{User-centric Sequential Models}}
The user-centric models mainly utilize the order information of consumed items to track users' interest drift.
Recurrent neural network (RNN)-based model \cite{beutel2018latent} naturally handles the sequential nature of consumed items.
Similarly, convolution neural network (CNN)-based models \cite{tang2018personalized,ma2019hierarchical} treat the consecutive items as an image with the convolution operation to compute the interaction among the items. SASRec \cite{kang2018self} applies the self-attention mechanism \cite{vaswani2017attention} to the recommender system to capture long-term dependency among consumed items compared to RNN and CNN. TiSASRec \cite{li2020time} extends SASRec \cite{kang2018self} by modeling the time interval between consecutive consumed items. 
Recently, LSAN \cite{li2021lightweight} captures local and global interactions among consumed items based on both CNN and self-attention modules.

Despite their success, these models do not explicitly consider whether users’ interest sustains beyond the training time as they depend on the next item prediction. \MD{} learns which items each user is likely to consume beyond the training time by predicting users' consumption in the recent period of the training time instead of the whole training time as in the next item prediction. 

\subsubsection{\textbf{Item-centric Sequential Models}}
Item-centric sequential recommendation is an under-studied topic. 
In contrast to the user-centric models, item-centric models capture users' general interest drift for each item by leveraging all users' consumption history for each item.
A precedent item-centric work \cite{wang2019modeling} considers the period of time after the last consumption of each item to predict a repetitive consumption of items in the future. 
CRIS \cite{hyun2020interest} learns whether users' general interest in items will sustain up to the future by predicting whether each item is consumed in the recent period of the training time. The recent item-centric model \cite{hyun2020interest} shows better recommendation accuracy than the user-centric models.

However, as these models only learn non-personalized interest drift, they tend to recommend generally-consumed items without considering each user's taste, e.g., \textit{vegan} or \textit{non-vegan}. \MD{} addresses the problem by predicting each user's consumption to capture the personalized interest sustainability instead of predicting all users' consumption for each item as in the item-centric models.

\begin{figure}[t]
    \centering
    \begin{subfigure}[b]{0.39\linewidth}
         \centering
         \includegraphics[height=4cm,width=\textwidth]{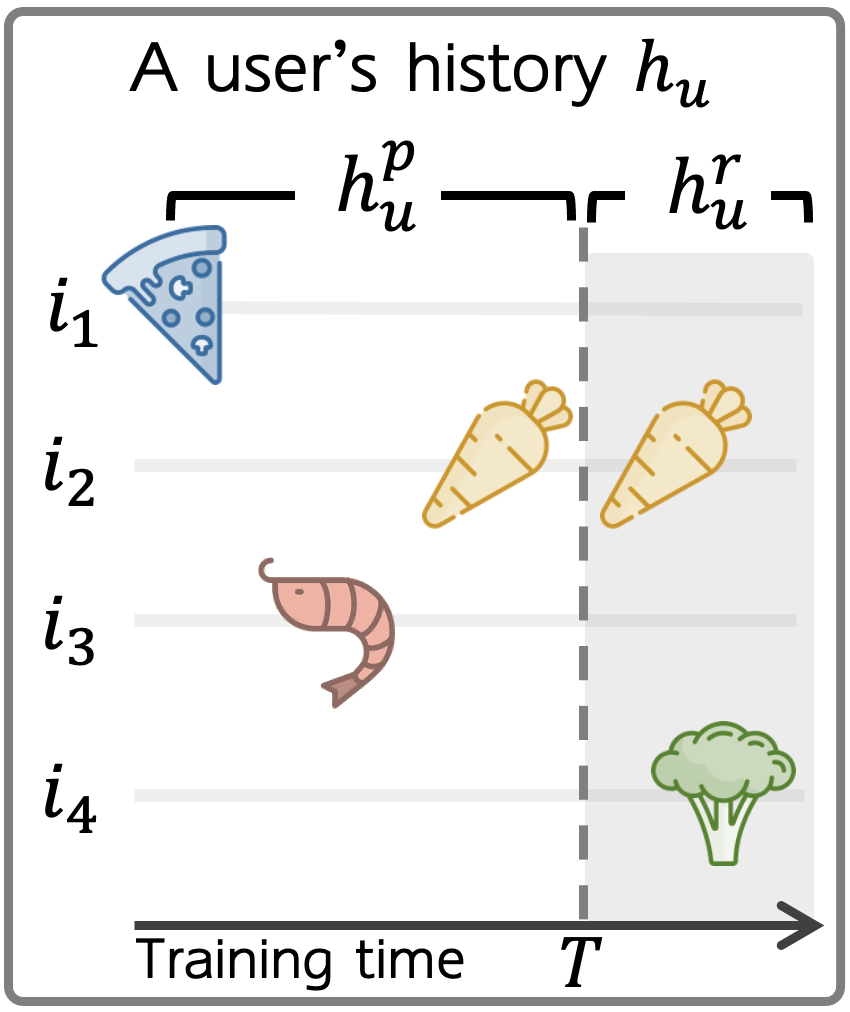}
         \caption{Data split}
         \label{fig:pisp_data}
     \end{subfigure}
     \begin{subfigure}[b]{0.6\linewidth}
         \centering
         \includegraphics[height=4cm,width=\textwidth]{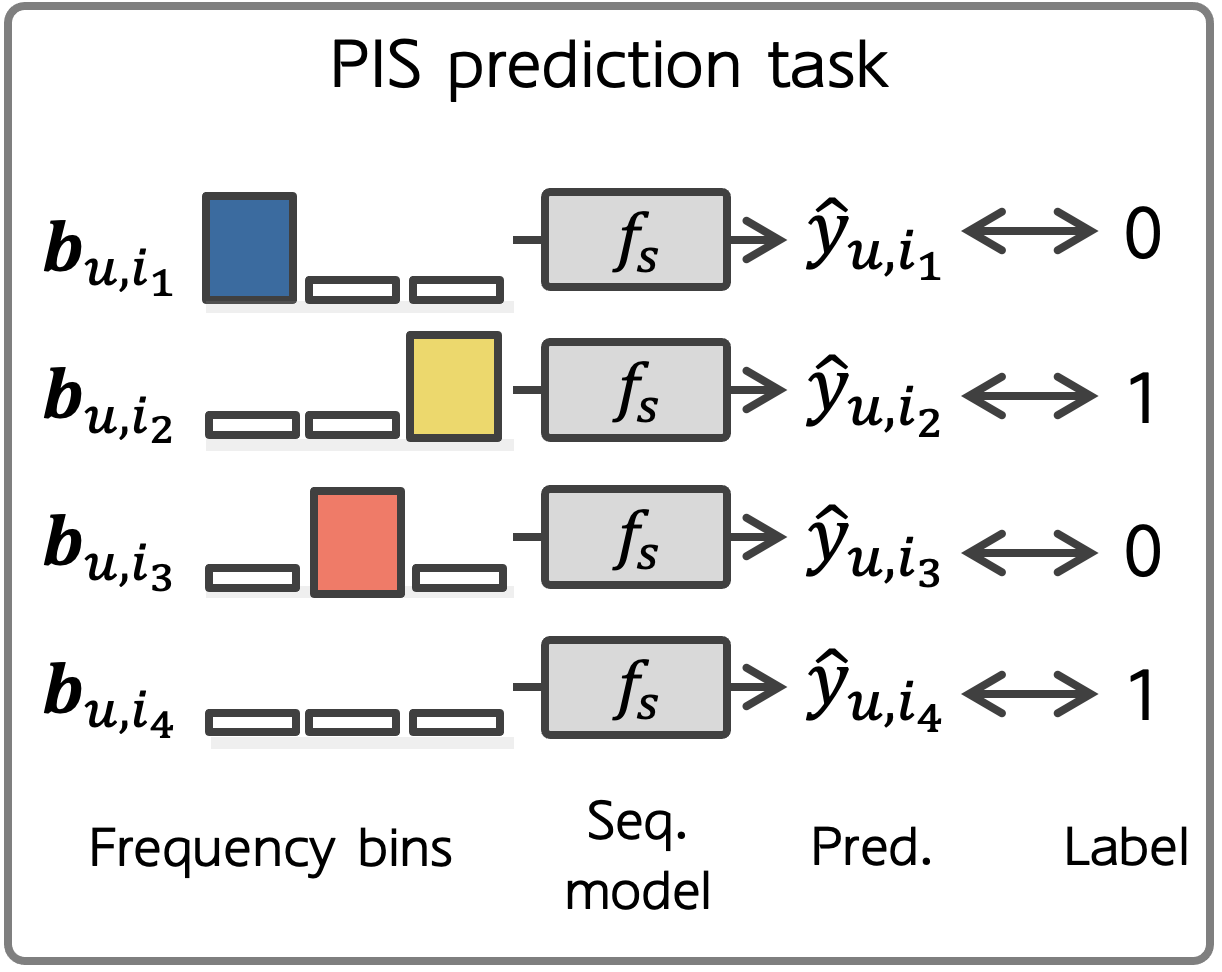}
         \caption{Model training}
         \label{fig:pisp_model}
     \end{subfigure}
    \caption{Illustration of a proposed prediction task.}
    \label{fig:pisp}
\end{figure}

\section{PERIS: Proposed Methodology}
We describe the problem (\cref{sec:probform}) and a new task to predict whether each user's interest in items sustains beyond the training time (\cref{sec:pis}). However, due to users' sparse consumption history, it is non-trivial to successfully perform the task based only on users' inherent consumption history. To this end, we propose simple yet effective intrinsic (\cref{sec:in}) and extrinsic (\cref{sec:ex}) schemes to supplement users' sparse consumption history. In addition, we complement a label noise of the newly-introduced task by adopting conventional preference learning (\cref{sec:pref}). We lastly describe the training loss and inference score (\cref{sec:loss}).

\subsection{Problem Formulation}
\label{sec:probform}
Let $ \mathcal{D} = \{(u, i, t)| \text{ user } u \text{ interacted with item } i \text{ at time } t\} $ be the training data and $\mathcal{U}$ and $\mathcal{I}$ are the set of users and items.
As input, a model takes user $u$ and the user's consumption history, i.e., $h_u = \{(i, t) | \text{ a user } u \text{ interacted with item } i \text{ at time } t\}$. 
In this work, recommender systems suggest top-\textit{K} items for users. 

\subsection{Personalized Interest Sustainability}
\label{sec:pis}

To overcome the limitations of the user- and item-centric models, we propose a task that requires to predict which items each user consumes in the recent period of the training time. The recommender system trained under this task can learn the personalized interest by predicting each user's consumption as in the user-centric models. In addition, the task requires to predict users' consumption occurred within the recent period of the training time. Thus, the model can focus on users' recent interest like the item-centric models, resulting in accurate recommendation thanks to the short period of time between the test time and the time at which users' last interest is captured. %

\subsubsection{\textbf{PIS Prediction Task}}
The goal of this task is to predict the personalized interest sustainability (PIS) defined as whether each user's interest in items is likely to sustain up to the future. 
We treat the recent period of the training time as the future, and control the length of the recent period based on a predefined time $T$, which is a tunable parameter. 
Specifically, as shown in Figure \ref{fig:pisp_data}, we divide a user's consumption history $h_u$ into the past and recent parts based on the predefined time $T$:
\begin{equation*}
    h_u = h_u^p \, \cup  \, h_u^r 
\end{equation*}
\begin{equation*}
    h_u^p = \{(i,t)| (i,t) \in h_u, t < T\}
\end{equation*}
\begin{equation*}
    h_u^r = \{(i,t)| (i,t) \in h_u, t \geq T\}
\end{equation*}
where $h_u^p$ and $h_u^r$ are the past and recent parts of each user's consumption history $h_u$.
Given the divided consumption history, the proposed task requires to predict which items each user $u$ consumes in the recent period, i.e., $\{i| i \in h_u^r\}$, based on the user's previous history $h_u^p$.
Formally, we define the label representing whether item $i$ is consumed by user $u$ in the recent period of the training time: 
\begin{equation}
    y_{u,i} = \mathbbb{1}[i \in h_u^r]
    \label{eqn:label}
\end{equation}
where $y_{u,i}$ is a binary label and $\mathbbb{1} \rightarrow{\{0, 1\}}$ is the indicator function.

\begin{figure}[t]
    \centering
    \begin{subfigure}[b]{0.6\linewidth}
         \centering
         \includegraphics[width=\textwidth]{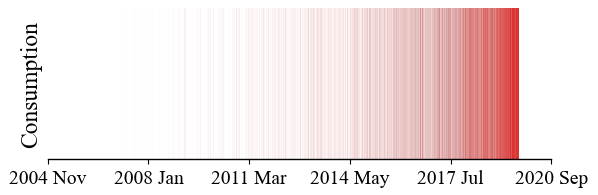}
         \caption{$y_{u,i}=1$}
         \label{fig:pattern1}
     \end{subfigure}
     
     \begin{subfigure}[b]{0.6\linewidth}
         \centering
         \includegraphics[width=\textwidth]{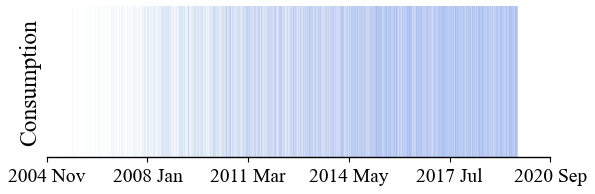}
         \caption{$y_{u,i}=0$}
         \label{fig:pattern0}
     \end{subfigure}
     \caption{Temporal consumption patterns belonging to each label. Each vertical line and color intensity denote the consumption time and the number of consumption, respectively.}
     \label{fig:pattern}
\end{figure}

\subsubsection{\textbf{Features for PIS Prediction}}
As the feature for performing this task, we consider users' temporal consumption pattern, i.e., the times at which a user consumed an item.
We first analyze how users' temporal consumption pattern belonging to each label is discriminative on Yelp data (Figure \ref{fig:pattern}). We overlap each user $u$'s the consumption time for each consumed item $i$ within the past part $h_u^p$, i.e., the period for defining features, according to whether user $u$ consumes item $i$ in the recent period (Figure \ref{fig:pattern1}) or not (Figure \ref{fig:pattern0}). 
Based on the analysis, we observe that the labels are discriminative based on the temporal consumption patterns, i.e., the more recently consumed, the more likely it will be consumed in the future, i.e., $y_{u,i}=1$ (Figure \ref{fig:pattern1}), whereas the other case does not show such clear consumption pattern, i.e., $y_{u,i}=0$ (Figure \ref{fig:pattern0}).

Thus, we utilize user $u$'s temporal consumption pattern for item $i$ as input feature as follows:
\begin{equation*}
C_{u,i}=\{t| (i,t) \in h_u^p\}
\end{equation*}
where $C_{u,i}$ is a set of user $u$'s consumption times for item $i$.
We then discretize the consumption times $C_{u,i}$ into sequential frequency bins (Figure \ref{fig:pisp_model}) with an equal width $w$ to capture the temporal dynamics of consumption. We split the whole period of time before the predefined time $T$, which is the period for defining the features, into a set of time intervals:
\begin{equation*}
    V_n=[\min(L) + (n-1)\cdot w, \,\, \min(L) + n \cdot w], \forall n \in\{1,...,N\}
\end{equation*}
where $V_n$ is a time interval for $n$-th frequency bin , $w$ is the bin width (e.g., one month), $L \in \{t|t\in \mathcal{D}, t < T\}$ is a set of consumption times in the whole period for defining features, $N$ is the number of time bins calculated by $N = \ceil{(\max(L) - \min(L))/w}$.
We then assign each consumption time $t$ into time bins based on corresponding time intervals as follows:
\begin{equation}
    b_{u,i}^n = \sum_{t\in C_{u,i}} \mathbbb{1}[t \in V_n], \quad \forall n \in \{1,...,N\}
    \label{eqn:bin_assign}
\end{equation}
where $b_{u,i}^n \in \mathbb{R}$ is $n$-th frequency bin. From the definition, we can obtain a sequence of the frequency bins, i.e., $\textbf{b}_{u,i} \in \mathbb{R}^N$. 

\subsubsection{\textbf{Training Objective}}
Given the feature, we predict the label $y_{u,i}$ defined as whether user $u$ consumes item $i$ in the recent period with a prediction model such that:
\begin{equation*}
     \hat{y}_{u,i} = f_s(\textbf{b}_{u,i}) 
\end{equation*}
where $\hat{y}_{u,i}$ is the prediction score and $f_s$ is a prediction model that predicts the label $y_{u,i}$ from the sequential feature $\textbf{b}_{u,i}$. The detailed architecture is provided in the following section (\cref{sec:detail}).

We then train the prediction model $f_s$ under the following loss:
\begin{equation}
     \mathcal{L} = \frac{1}{|\mathcal{U}|} \sum_{u\in \mathcal{U}} \sum_{i\in h_u} (y_{u,i} - \hat{y}_{u,i})^2
     \label{eqn:l2loss}
\end{equation}
We note that, for each user, we set the candidate items for training as the items consumed by each user, i.e., $\{i| i\in h_u\}$, instead of all the items. Thus, we can train the model more efficiently by focusing on items in which users are interested than exhaustively predicting users' consumption for every item.

After training, we can predict users' interest beyond the training time by expanding the frequency bins up to the whole training time and feeding the expanded frequency bins to the trained model $f_s$:
\begin{equation*}
    \bar{V}_n=[\min(\bar{L}) + (n-1)\cdot w, \,\, \min(\bar{L}) + n \cdot w], \forall n \in\{1,...,M\}
\end{equation*}
where $\bar{V}_n$ is the expanded time interval up to the end of the training time, $\bar{L} \in \{t|t\in D\}$ is a set of consumption times in the training data $\mathcal{D}$ and $M \ge N$ where $M = \ceil{(\max(\bar{L}) - \min(\bar{L}))/w}$. We can obtain the expanded frequency bins ($\textbf{b}_{u,i} \in \mathbb{R}^M$) based on Equation \ref{eqn:bin_assign} with the new time interval $\bar{V}_n, \forall n \in \{1,...,M\}$.

\subsubsection{\textbf{Details of Prediction Model}}
\label{sec:detail}
We here provide the details of the prediction model $f_s$. In this work, we design a prototype-based classifier similar to \cite{mettes2019hyperspherical} as we observe that it produces higher accuracy than other alternatives such as the multi-layer perceptron thanks to the small number of parameters to avoid the overfitting:
\begin{equation*}
    f_s(\textbf{b}_{u,i}) = 1 - d(\mathcal{C}, \textbf{h}_{u,i})
\end{equation*}
\begin{equation*}
    \textbf{h}_{u,i}  = \overrightarrow{LSTM}(\textbf{b}_{u,i}) 
\end{equation*}
where $\mathcal{C} \in \mathbb{R}^k$ is a trainable parameter representing a positive class (i.e., $y_{u,i}=1$), called prototype, and $d$ is the euclidean distance. 
To capture the sequential dynamics of the feature $\textbf{b}_{u,i}$, we use Long Short-Term Memory (LSTM) \cite{hochreiter1997long}, and the prediction model $f_s$ predicts the label based on the last hidden representation $\textbf{h}_{u,i}\in \mathbb{R}^k$ from LSTM. 
Thus, the model $f_s$ classifies the frequency bins $\textbf{b}_{u,i}$ (i.e., the feature) as the positive label, i.e., $\hat{y}_{u,i}=1$, if its hidden representation $\textbf{h}_{u,i}$ is close to the prototype $\mathcal{C}$, and vice versa.

A remaining issue is that users consume items only few times in some domain, which degrades the prediction accuracy due to the sparse consumption history, e.g., zero values in most frequency bins $\textbf{b}_{u,i}$ as shown in Figure \ref{fig:pisp_model}. To address the sparsity issue, we devise intrinsic and extrinsic supplementation schemes.

\begin{figure}[t]
  \includegraphics[width=0.99\linewidth]{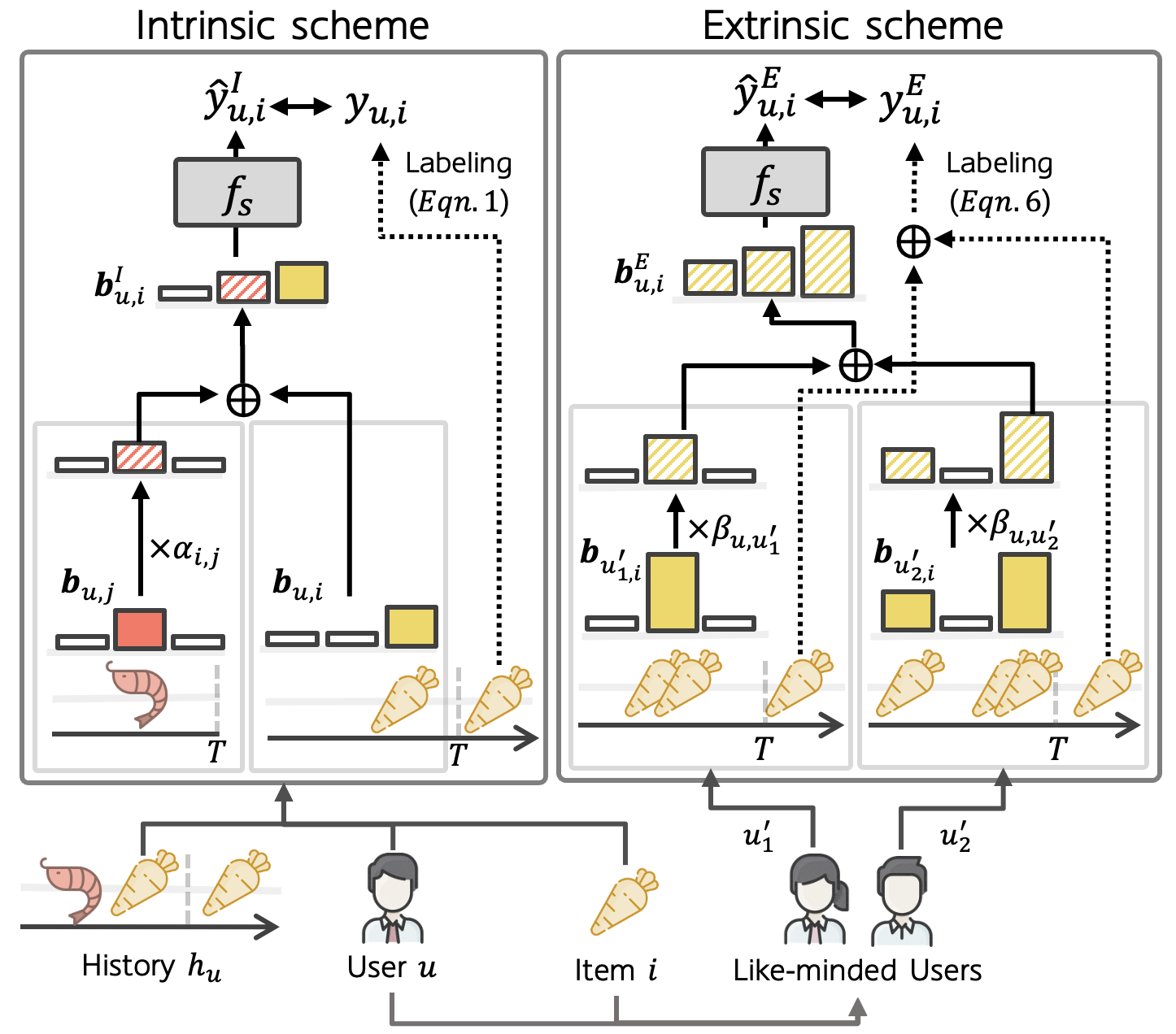}
  \caption{Example of intrinsic and extrinsic supplementation schemes. Strips depict supplemented frequency bins. $\oplus$ denotes the element-wise addition. 
  }
  \label{fig:schemes}
\end{figure}

\subsection{Intrinsic Supplementation Scheme}
\label{sec:in}
We first propose an intrinsic scheme (Figure \ref{fig:schemes}) to alleviate the data sparsity by augmenting each user's consumption history for an item based on \textit{other items consumed by the user}. 
Its underlying idea is that a user’s interest in an item (e.g., \textit{espresso}) is assumed to sustain if the user recently consumes a similar item (e.g., \textit{cappuccino}).
 
Formally, the goal of this intrinsic scheme is to augment user $u$'s feature for item $i$, i.e., $\textbf{b}_{u,i}$, by utilizing features of other items consumed by user $u$ based on the item-wise similarity: 
\begin{equation*}
    \textbf{b}_{u,i}^I = \textbf{b}_{u,i} + \sum_{j \in h_u^p \setminus{\{i\}}}\alpha_{i,j} \cdot \textbf{b}_{u,j}
\end{equation*}
\begin{equation}
    \alpha_{i,j} = \texttt{sim}(\textbf{v}_i, \textbf{v}_j)
    \label{eqn:sim}
\end{equation}
where $\textbf{b}_{u,i}^I\in \mathbb{R}^N$ denotes user $u$'s augmented feature for item $i$ by aggregating the features of user $u$'s other consumed items. In addition, $\mathtt{sim}$ is the normalized cosine similarity (i.e., $\mathtt{sim}(\cdot, \cdot)=(\mathtt{cos}(\cdot,\cdot)+1)/2 + \tau$ ) with a tunable scaling parameter $\tau \in \mathbb{R}$, and $\textbf{v}_i \in \mathbb{R}^k$ is an embedding vector for item $i$. 
Thus, this intrinsic scheme supplements each user's consumption history of a target item by referring to the consumption history of other highly-relevant items consumed by the user.
We then make the prediction as follows:
\begin{equation}
    \hat{y}_{u,i}^I = f_s(\textbf{b}_{u,i}^I + \textbf{e}_{u,i})
    \label{eqn:pred}
\end{equation}
where $\hat{y}_{u,i}^I$ is the prediction based on the feature from the intrinsic scheme, i.e., $\textbf{b}_{u,i}^I$. In addition, we include a user-item joint representation (i.e., $\textbf{e}_{u,i}=\textbf{u}_u+\textbf{v}_i$) as an additional input to provide a model the information of a target user $u$ and item $i$ where $\textbf{u}_u \in \mathbb{R}^k$ is an embedding vector for user $u$. Note that we adopt warm-up steps to prepare the embedding vectors as they are randomly initialized, the details are provided in the following section (\cref{sec:loss}).

\subsection{Extrinsic Supplementation Scheme}
\label{sec:ex}
We then propose an extrinsic scheme (Figure \ref{fig:schemes}) to supplement the consumption history of each user (e.g., \textit{vegan}) by referring to the consumption history of \textit{other like-minded users} (e.g., \textit{other vegans}). In a nutshell, the goal is to infer a target user's interest in items through the like-minded users' interest in those items. 

We first aggregate like-minded users' features for target item $i$:
\begin{equation*}
    \textbf{b}_{u,i}^E= \sum_{\textbf{b}_{u',i} \in B_{u,i}} \beta_{u, u'} \cdot \textbf{b}_{u',i}
\end{equation*}
where $\textbf{b}_{u,i}^E$ is the aggregated features of like-minded users of a target user $u$ for a target item $i$, and other users are denoted by $u'$. $B_{u,i}$ is a set of features of other users who consume the target item $i$ except user $u$ and $\beta_{u, u'}$ is the similarity between a target user $u$ and other users $u'$, which are formalized as follows:
\begin{equation*}
    B_{u,i} = \{ \textbf{b}_{u',i}|| u' \in \mathcal{U} \setminus{\{u\}}, \, i \in h_{u'}^p, \}
\end{equation*}
\begin{equation*}
    \beta_{u,u'} = \texttt{sim}(\textbf{u}_u, \textbf{u}_{u'})
\end{equation*}
The similarity function $\mathtt{sim}$ is identical to Equation \ref{eqn:sim}, which is defined with the cosine similarity. 

We also set the label of the like-minded users' consumption of a target item $i$ based on Equation \ref{eqn:label} such that:
\begin{equation}
    y_{u,i}^E = \mathbbb{1}[\sum_{u'} \beta_{u, u'} \cdot y_{u',i} \ge 1]
\end{equation}
where the aggregated label $y_{u,i}^E$ indicates that the like-minded users consume the target item $i$ if their weighted consumption is greater than one consumption, i.e., item $i$ is consumed at least one time. 
Given the extrinsic feature $\textbf{b}_{u,i}^E$ and label $y_{u,i}^E$, we perform the prediction based on Equation \ref{eqn:l2loss}.

\subsection{Preference Learning}
\label{sec:pref}
The PIS prediction task enables the model to predict which items each user will consume in the future, but the label $y_{u,i}$ can be noisy because a user can still prefer an item even though the user does not consume the item in the recent period, e.g., a user has a long consumption period. To complement the label noise, we train \MD{} along with conventional preference learning, which trains the model with ground-truth labels, i.e., consumed items. 
Among existing methods, we adopt a metric learning-based method with a prototype \cite{movshovitz2017no, hyun2020interest} as it empirically shows better results than conventional methods such as BPR \cite{mnih2008probabilistic}.

We first obtain a joint representation for a user-item pair:
\begin{equation*}
    \textbf{e}_{u,i} = \textbf{u}_u + \textbf{v}_i
\end{equation*}
where $\textbf{e}_{u,i} \in \mathbb{R}^k$ is the joint representation. Given the joint representation, the preference learning is formulated as follows:
\begin{equation*}
    \mathcal{L}_{P} = \sum_{u \in \mathcal{U}} \sum_{i^+ \in h_u, i^- \notin h_u} [d(\mathcal{P}, \textbf{e}_{u,i^+}) - d(\mathcal{P}, \textbf{e}_{u,i^-})+m]_+
\end{equation*}
where $d$ is the euclidean distance, $i^+$ is a positive item consumed by users, $i^-$ is a negative item not consumed by users, $m \in \mathbb{R}$ is a margin between the positive and negative interactions (i.e., $\textbf{e}_{u,i^+}$ and $\textbf{e}_{u,i^-}$, respectively) from a prototype $\mathcal{P} \in \mathbb{R}^k$ that is a trainable parameter (i.e., a prototype), and $[\cdot]_+$ denotes $\max(\cdot,0)$. 
Thus, the more likely user $u$ is to consume item $i$, the closer the joint representation $\textbf{e}_{u,i}$ is to the prototype $\mathcal{P}$, and vice versa.

\subsection{Model Training and Inference}
\label{sec:loss}
The final loss is the combination of the training objectives:
\begin{equation}
    \mathcal{L}_F = \lambda \bigl \{\mu \mathcal{L}_{I} + (1-\mu)\mathcal{L}_{E}\bigr\} + (1-\lambda) \mathcal{L}_{P} 
    \label{eqn:final_loss}
\end{equation}
where $\lambda$ and  $\mu$ are tunable coefficients to control the balance among the losses, $\mathcal{L}_I$ is a loss computed with the intrinsic scheme (i.e., $\mathcal{L}_I = \frac{1}{|\mathcal{U}|} \sum_{u\in \mathcal{U}} \sum_{i\in h_u} (y_{u,i} - \hat{y}_{u,i}^I)^2$), and $\mathcal{L}_E$ is a loss computed with the extrinsic scheme (i.e., $\mathcal{L}_E = \frac{1}{|\mathcal{U}|} \sum_{u\in \mathcal{U}} \sum_{i\in h_u} (y_{u,i}^E - \hat{y}_{u,i}^E)^2$).
We also adopt warm-up steps to train \MD{} with only the loss of the preference learning $\mathcal{L}_{P}$ since the similarity used by intrinsic and extrinsic schemes is computed based on user and item embedding vectors, which are otherwise randomly initialized. After the warm-up steps, we train \MD{} with the final loss $\mathcal{L}_F$.

We then compute the recommendation score as follows:
\begin{equation}
    r_{u,i} = \lambda \bigl \{\mu \hat{y}_{u,i}^I + (1-\mu)\hat{y}_{u,i}^E\bigr\} + (1-\lambda) \hat{y}_{u,i}^P
    \label{eqn:score}
\end{equation}
where $r_{u,i}\in \mathbb{R}$ is the final recommendation score, $\hat{y}_{u,i}^I$ and $\hat{y}_{u,i}^E$ are user $u$'s interest in item $i$ from the intrinsic and extrinsic schemes, and $\hat{y}_{u,i}^P = -d(\mathcal{P}, \textbf{e}_{u,i})$ is the interest from the preference learning.  

\smallskip
\noindent\textbf{Comparison to Item-Centric Model.} The recent item-centric model \cite{hyun2020interest} learns the non-personalized interest sustainability (NIS). However, \MD{} advances the interest sustainability in three aspects. 
First, we formulate the PIS, which can better track users' interest drift thanks to the consideration of each user's interest compared to all users' general interest. Moreover, we devise simple yet effective schemes to supplement users' sparse consumption history, and observe that, without the schemes, the PIS prediction task harms the recommendation accuracy due to users' sparse consumption history.
Second, \MD{} is an end-to-end method, whereas the item-centric model consists of two independent training steps. The item-centric approach first obtains the NIS before training a recommender system, then trains the recommender system while fixing NIS during the training. Therefore, the NIS can be sub-optimal with respect to the recommendation as the NIS is obtained without considering the recommendation performance. In contrast, we train \MD{} by simultaneously performing both tasks, i.e., the PIS prediction task and preference learning, which enables to learn  the PIS with considering the recommendation performance.
Third, \MD{} can easily update each user's PIS by adding newly-consumed items to the consumption history, but the item-centric model requires to re-train the model to update users' new consumption as the model depends on the fixed NIS. Therefore, \MD{} substantially enhances the recommendation accuracy as we will see in the experiments.

\section{Experiments}

\subsection{Experimental Settings}

\subsubsection{\textbf{Datasets}}
We evaluate \MD{} compared to 10 baseline recommender systems on 11 real-world datasets. Amazon datasets\footnote{\url{jmcauley.ucsd.edu/data/amazon}} contain users' consumption history in the product shopping domains and have been commonly used benchmark datasets to evaluate recommender systems \cite{ma2019hierarchical,mao2021simplex}. We use the datasets in 9 categories (i.e., from Cell phones to CDs in Table \ref{tab:data}) to consider the scenario of the recommendation in diverse domains. Yelp dataset\footnote{\url{www.yelp.com/dataset}} contains users' consumption history in various services, e.g., hotels and restaurants.
Google dataset\footnote{\url{cseweb.ucsd.edu/~jmcauley/datasets.html}} contains users' interaction with businesses from Google Maps.
We note that the published Yelp dataset is intentionally filtered by the Yelp system\footnote{\url{www.yelp.com/dataset/documentation/faq}}, making the data highly dense with respect to items, i.e., 28.1 interactions for each item on average. Thus, we make the Yelp dataset to have a similar statistic to raw data, i.e., the Amazon and Google datasets, by randomly dropping the interactions associated with items.  
On all the datasets, we filter out noisy data by dropping users who consumed less than 10 items \cite{hsieh2017collaborative,tang2018personalized,li2020symmetric}. 
We also exclude cold-start users/items, i.e., users/items that do not appear in the training time, as addressing the cold-start problem is a separate issue \cite{volkovs2017content,elkahky2015multi} and thus out of scope of this work.
Table \ref{tab:data} reports the statistics of the datasets. 
\subsubsection{\textbf{Baselines}}
We consider the following baseline models.

1) General recommender system
\begin{itemize}
    \item[$-$] \textbf{BPR} \cite{rendle2012bpr} learns users' preference based on a pair-wise ranking loss with the positive and negative items.
    \item[$-$] \textbf{CML} \cite{hsieh2017collaborative}  addresses the triangle inequality issue of the inner product with a distance-based metric learning.
    \item[$-$] \textbf{SML} \cite{li2020symmetric} is an extension of CML that adopts a symmetric learning mechanism and adaptive margin parameters.
    \item[$-$] \textbf{SimpleX} (SimX) \cite{mao2021simplex} is the state-of-the-art general model that uses cosine similarity and contrastive learning.
\end{itemize}

2) User-centric sequential recommender system
\begin{itemize}
    \item[$-$] \textbf{Caser} \cite{tang2018personalized} adopts convolutional neural network (CNN) to capture local context in each user's consumption history.
    \item[$-$] \textbf{SASRec} (SSR) \cite{kang2018self} utilizes self-attention to capture long-term dependency between items in consumption history. 
    \item[$-$] \textbf{TiSASRec} (TSSR) \cite{li2020time} extends SASRec by modeling time intervals between items in users' consumption history.
    \item[$-$] \textbf{HGN} \cite{ma2019hierarchical} captures users' long- and short-term interest based on a hierarchical gating network.
    \item[$-$] \textbf{LSAN} \cite{li2021lightweight} is the state-of-the-art user-centric sequential model that combines CNN and self-attention to capture both local and global contexts.
\end{itemize}

3) Item-centric sequential recommender system
\begin{itemize}
    \item[$-$] \textbf{CRIS} \cite{hyun2020interest} is the state-of-the-art item-centric sequential model that captures users' general interest drift from all user' consumption history for each item.
\end{itemize}

\begin{table}[]
\renewcommand{\arraystretch}{1.2}
\caption{Data statistics. Int./user (item) denotes the averaged number of the interactions associated with users (items).}
\resizebox{\linewidth}{!}{
\begin{tabular}{lcccccc}
\toprule
\multirow{2}{*}{Data}	    & \multirowcell{2}{\# users \\ ($|\mathcal{U}|$)}	& \multirowcell{2}{\# items \\ ($|\mathcal{I}|$)} & \multirowcell{2}{\# data \\ ($|\mathcal{D}|$)} & \multirowcell{2}{Int./\\user} & \multirowcell{2}{Int./\\item}	& \multirow{2}{*}{Time span} \\ \\ \midrule

Cell phones	    &8,192	&47,671	&118,323	&14.44	&2.48	& 2000.10-2014.05 \\
Digital music	        &6,062	&65,094	&127,484	&21.03	&1.96	& 1998.05-2014.05 \\
Tools	    &8,971	&61,271	&150,780	&16.81	&2.46	& 1999.11-2014.05 \\
Grocery 	&8,215	&54,452	&168,933	&20.56	&3.10	& 2000.08-2014.05 \\
Toys	    &12,636	&99,051	&230,473	&18.24	&2.33	& 1999.10-2014.05 \\
Health	    &14,149	&68,810	&252,356	&17.84	&3.67	& 2000.12-2014.05 \\
Sports	    &16,959	&99,927	&276,214	&16.29	&2.76	& 2000.07-2014.05 \\
Clothing	&35,824	&315,818	&578,135	&16.14	&1.83	& 2000.11-2014.05 \\
CDs	        &40,339	&330,179	&1,278,176	&31.69	&3.87	& 1997.11-2014.05 \\
Yelp	    &18,284	&83,871	&384,330	&21.02	&4.58	& 2004.10-2020.11 \\
Google	    &125,341	&1,552,812	&3,066,438	&24.46	&1.97	& 1990.12-2014.01 \\ \bottomrule
\end{tabular}
}
\label{tab:data}
\end{table}

\subsubsection{\textbf{Evaluation Protocol}}
We split the datasets into training, validation, and test data based on the interaction times. Specifically, we set the last one month, i.e., 30 days, as the test data by following \cite{hyun2020interest}. 
We then set the last month of the remaining data as the validation data, and the final remaining data is used as the training data.
As for the performance metrics, we use two widely-used metrics: hit ratio (HR) and normalized discounted cumulative gain (nDCG). The HR measures whether the recommendations from models include items that users consumed. Moreover, nDCG considers the position of the consumed items in the recommendation list, i.e., the higher the position is, the higher the score. We consider top-\textit{K} recommendation, and thus report the HR@\textit{K} (H@\textit{K}) and nDCG@\textit{K} (N@\textit{K}). 
Following a commonly-used evaluation protocol \cite{li2020time,hyun2020interest,cho2021learning}, we measure the metrics for each consumed item compared to 100 randomly-sampled items which are not consumed by a target user. We run models 5 times and report the mean of the metrics \cite{ma2019hierarchical,cho2021unsupervised}.

\begin{table*}[]
\small
\centering
\caption{Comparison of recommendation accuracy. The results are in percentage without `\%' for brevity. $\Delta_G$ and $\Delta_S$ denote the relative improvement of \MD{} over the best result from general and sequential model. I-SRS: Item-centric sequential RS. The results of \MD{} are  statistically significant compared to the best baseline model for each dataset with \textit{p} < 0.001 from the t-test.}
\renewcommand{\arraystretch}{.79}
\resizebox{0.75\linewidth}{!}{
\begin{tabular}{llccccccccccccc}
\toprule
\multicolumn{2}{c}{Setting}                           & \multicolumn{4}{c}{General RS} & \multicolumn{5}{c}{User-centric Sequential RS}   & \multicolumn{1}{c}{I-SRS}    & \multicolumn{3}{c}{Proposed RS} \\ \cmidrule(lr){1-2} \cmidrule(lr){3-6} \cmidrule(lr){7-11} \cmidrule(lr){12-12} \cmidrule(lr){13-15}
Dataset                   & Metric & BPR   & CML   & SML   & SimX & SSR & TSSR & Caser & HGN   & LSAN  & CRIS  & PERIS & $\Delta_G$   & $\Delta_S$   \\ \midrule
\multirowcell{4}{Cell\\Phones}     & H@5  & 48.17 & 48.91 & 50.85       & 52.89       & 45.79 & 50.15 & 43.78 & 51.95 & 54.21       & {\ul 56.07} & \textbf{63.68} & 20.4\% & 13.6\% \\
         & H@10 & 59.61 & 60.68 & 63.57       & 63.84       & 59.32 & 63.31 & 56.73 & 63.63 & 65.55       & {\ul 68.38} & \textbf{76.28} & 19.5\% & 11.6\% \\
         & N@5  & 35.34 & 36.50 & 38.02       & 39.57       & 32.44 & 36.36 & 32.28 & 39.63 & 40.79       & {\ul 43.19} & \textbf{48.74} & 23.2\% & 12.9\% \\
         & N@10 & 39.06 & 40.32 & 42.16       & 43.09       & 36.84 & 40.62 & 36.46 & 43.42 & 44.47       & {\ul 47.20} & \textbf{52.84} & 22.6\% & 11.9\% \\ \midrule
\multirowcell{4}{Digital\\Music}       & H@5  & 35.21 & 34.49 & {\ul 38.46} & 37.22       & 24.10 & 25.00 & 25.56 & 25.38 & 34.83       & 27.78       & \textbf{47.31} & 23.0\% & 35.8\% \\
         & H@10 & 44.53 & 44.49 & {\ul 50.21} & 46.54       & 31.97 & 36.37 & 36.28 & 34.53 & 44.79       & 37.78       & \textbf{58.50} & 16.5\% & 30.6\% \\
         & N@5  & 25.71 & 25.30 & {\ul 28.32} & 27.98       & 16.53 & 16.99 & 18.56 & 18.64 & 25.46       & 19.38       & \textbf{35.31} & 24.7\% & 38.7\% \\
         & N@10 & 28.73 & 28.53 & {\ul 32.12} & 30.98       & 19.06 & 20.52 & 22.05 & 21.58 & 28.70       & 22.62       & \textbf{38.93} & 21.2\% & 35.6\% \\ \midrule
\multirowcell{4}{Tools}    & H@5  & 36.55 & 36.53 & 37.78       & {\ul 39.83} & 34.79 & 32.96 & 30.70 & 34.92 & 37.95       & 35.61       & \textbf{43.64} & 9.6\%  & 15.0\% \\
         & H@10 & 48.89 & 47.86 & 50.50       & {\ul 52.69} & 48.03 & 47.45 & 44.95 & 47.45 & 48.93       & 49.77       & \textbf{57.15} & 8.5\%  & 14.8\% \\
         & N@5  & 25.67 & 25.80 & 26.39       & {\ul 28.13} & 23.48 & 22.65 & 20.66 & 24.35 & 26.40       & 24.19       & \textbf{31.35} & 11.4\% & 18.8\% \\
         & N@10 & 29.66 & 29.46 & 30.50       & {\ul 32.30} & 27.78 & 27.31 & 25.27 & 28.41 & 29.94       & 28.75       & \textbf{35.73} & 10.6\% & 19.3\% \\ \midrule
\multirowcell{4}{Grocery}  & H@5  & 46.85 & 46.27 & 47.61       & 46.78       & 45.36 & 46.12 & 43.46 & 45.24 & 46.27       & {\ul 49.50} & \textbf{51.86} & 8.9\%  & 4.8\%  \\
         & H@10 & 56.93 & 56.64 & 58.32       & 57.98       & 57.10 & 57.19 & 55.36 & 56.28 & 57.26       & {\ul 60.60} & \textbf{62.91} & 7.9\%  & 3.8\%  \\
         & N@5  & 33.91 & 33.91 & 34.86       & 34.99       & 33.98 & 34.00 & 30.03 & 34.41 & 34.33       & {\ul 37.13} & \textbf{40.55} & 15.9\% & 9.2\%  \\
         & N@10 & 37.17 & 37.27 & 38.32       & 38.61       & 37.78 & 37.56 & 33.88 & 37.98 & 37.88       & {\ul 40.73} & \textbf{44.13} & 14.3\% & 8.3\%  \\ \midrule
\multirowcell{4}{Toys}     & H@5  & 45.87 & 45.89 & 46.84       & 48.53       & 34.22 & 35.11 & 35.87 & 42.41 & 48.73       & {\ul 49.54} & \textbf{53.21} & 9.6\%  & 7.4\%  \\
         & H@10 & 57.08 & 57.28 & 58.73       & 61.74       & 48.73 & 52.02 & 48.61 & 55.28 & 61.46       & {\ul 62.63} & \textbf{66.61} & 7.9\%  & 6.4\%  \\
         & N@5  & 33.99 & 33.67 & 34.91       & 35.46       & 22.62 & 23.70 & 25.37 & 31.28 & 36.18       & {\ul 36.68} & \textbf{39.64} & 11.8\% & 8.1\%  \\
         & N@10 & 37.61 & 37.35 & 38.75       & 39.75       & 26.97 & 29.11 & 29.50 & 35.42 & 40.31       & {\ul 40.92} & \textbf{43.99} & 10.7\% & 7.5\%  \\ \midrule
\multirowcell{4}{Health}   & H@5  & 46.43 & 47.97 & 49.07       & 51.18       & 43.50 & 44.92 & 46.06 & 49.03 & 50.31       & {\ul 52.43} & \textbf{53.99} & 5.5\%  & 3.0\%  \\
         & H@10 & 60.20 & 60.35 & 61.57       & 62.78       & 57.93 & 58.60 & 56.20 & 59.99 & 61.84       & {\ul 64.92} & \textbf{66.38} & 5.7\%  & 2.2\%  \\
         & N@5  & 32.71 & 34.55 & 35.09       & 37.68       & 31.72 & 32.23 & 34.42 & 37.72 & 37.64       & {\ul 38.77} & \textbf{42.48} & 12.7\% & 9.6\%  \\
         & N@10 & 37.17 & 38.55 & 39.14       & 41.43       & 36.37 & 36.67 & 37.70 & 41.24 & 41.38       & {\ul 42.82} & \textbf{46.48} & 12.2\% & 8.5\%  \\ \midrule
\multirowcell{4}{Sports}   & H@5  & 47.90 & 48.73 & 49.18       & {\ul 50.85} & 39.59 & 39.81 & 40.78 & 46.82 & 48.89       & 49.81       & \textbf{54.30} & 6.8\%  & 9.0\%  \\
         & H@10 & 58.53 & 60.23 & 60.66       & {\ul 64.12} & 52.38 & 54.35 & 53.38 & 59.07 & 60.39       & 61.38       & \textbf{66.82} & 4.2\%  & 8.9\%  \\
         & N@5  & 35.95 & 36.33 & 36.69       & 37.00       & 27.43 & 28.26 & 29.50 & 34.56 & 36.78       & {\ul 37.52} & \textbf{40.62} & 9.8\%  & 8.3\%  \\
         & N@10 & 39.38 & 40.04 & 40.41       & {\ul 41.30} & 31.57 & 32.74 & 33.57 & 38.51 & 40.51       & 41.27       & \textbf{44.68} & 8.2\%  & 8.3\%  \\ \midrule
\multirowcell{4}{Clothing} & H@5  & 39.26 & 40.01 & 36.83       & {\ul 46.95} & 40.17 & 43.51 & 39.88 & 38.78 & 38.99       & 45.36       & \textbf{50.48} & 7.5\%  & 11.3\% \\
         & H@10 & 48.21 & 50.14 & 46.89       & {\ul 58.45} & 51.25 & 57.18 & 52.91 & 50.64 & 50.23       & 57.07       & \textbf{64.70} & 10.7\% & 13.2\% \\
         & N@5  & 30.06 & 29.97 & 27.06       & {\ul 35.18} & 28.15 & 31.51 & 28.31 & 27.79 & 28.47       & 33.43       & \textbf{36.31} & 3.2\%  & 8.6\%  \\
         & N@10 & 32.95 & 33.24 & 30.30       & {\ul 38.90} & 31.74 & 35.92 & 32.53 & 31.63 & 32.10       & 37.22       & \textbf{40.90} & 5.1\%  & 9.9\%  \\ \midrule
\multirowcell{4}{CDs}      & H@5  & 62.96 & 62.16 & 62.96       & 59.58       & 37.32 & 42.04 & 56.15 & 56.96 & 60.83       & {\ul 63.05} & \textbf{65.13} & 3.4\%  & 3.3\%  \\
         & H@10 & 75.53 & 74.83 & {\ul 75.57} & 71.28       & 50.29 & 53.68 & 67.87 & 68.44 & 73.56       & 73.90       & \textbf{76.25} & 0.9\%  & 3.2\%  \\
         & N@5  & 47.57 & 47.40 & 48.51       & 45.30       & 26.19 & 30.32 & 43.18 & 44.15 & 46.51       & {\ul 49.37} & \textbf{51.45} & 6.1\%  & 4.2\%  \\
         & N@10 & 51.65 & 51.51 & 52.63       & 49.12       & 30.07 & 34.12 & 46.98 & 47.87 & 50.64       & {\ul 52.88} & \textbf{55.06} & 4.6\%  & 4.1\%  \\ \midrule
\multirowcell{4}{Yelp}     & H@5  & 66.13 & 63.60 & 65.21       & 65.21       & 43.41 & 46.36 & 62.38 & 64.34 & {\ul 68.79} & 66.09       & \textbf{73.93} & 11.8\% & 7.5\%  \\
         & H@10 & 84.53 & 82.74 & 82.42       & 84.11       & 61.75 & 63.75 & 78.21 & 80.02 & {\ul 85.63} & 84.38       & \textbf{87.29} & 3.3\%  & 1.9\%  \\
         & N@5  & 47.66 & 45.98 & 47.19       & 46.50       & 28.59 & 31.93 & 45.20 & 46.60 & {\ul 50.59} & 46.90       & \textbf{55.06} & 15.5\% & 8.8\%  \\
         & N@10 & 53.66 & 52.20 & 52.78       & 52.65       & 34.55 & 37.58 & 50.37 & 51.73 & {\ul 56.06} & 52.84       & \textbf{59.45} & 10.8\% & 6.0\%  \\ \midrule
\multirowcell{4}{Google}   & H@5  & 60.38 & 74.40 & {\ul 75.83} & 73.01       & 33.09 & 37.76 & 42.63 & 67.31 & 60.46       & 74.01       & \textbf{81.01} & 6.8\%  & 9.5\%  \\
         & H@10 & 68.69 & 81.08 & {\ul 82.54} & 78.62       & 44.54 & 48.01 & 53.97 & 73.83 & 72.76       & 81.94       & \textbf{86.83} & 5.2\%  & 6.0\%  \\
         & N@5  & 49.43 & 63.27 & {\ul 64.67} & 61.49       & 23.23 & 28.03 & 31.30 & 57.65 & 46.83       & 61.88       & \textbf{70.33} & 8.8\%  & 13.7\% \\
         & N@10 & 52.12 & 65.44 & {\ul 66.85} & 63.33       & 26.93 & 31.35 & 34.97 & 59.76 & 50.81       & 64.47       & \textbf{72.23} & 8.0\%  & 12.0\% \\ \bottomrule
\end{tabular}}
\label{tab:perf}
\end{table*}

\subsubsection{\textbf{Implementation Details}}
For fair comparison, we implement \MD{} and the baseline models in a unified framework based on PyTorch library \cite{paszke2019pytorch}. We tune their hyperparameters by grid search. We tune a learning rate of Adam optimizer \cite{kingma2014adam} in $\{0.01, 0.005, 0.001\}$, the mini-batch size in $\{64, 128, 256, 512, 1024, 2048\}$, the dimension for the embeddings (i.e., $k$) in $\{16, 32, 64, 128\}$, and we initialize user and item embeddings in all models with Xavier initialization \cite{glorot2010understanding}. We also normalize the user and item embeddings of CML-based models including \MD{} into a unit sphere to alleviate the curse of dimensionality issue as suggested in \cite{bordes2013translating,hsieh2017collaborative}. We tune other hyperparameters of the baseline models as reported in their literature. 
In the case of \MD{}, $\lambda$ and $\mu$ are tuned in $\{0.1, 0.3, 0.5, 0.7, 1\}$, the bin width $w$ in $\{4, 8, 12\}$ weeks, the scaling parameter $\tau$ in $\{0, 0.3, 0.5, 0.7\}$. We tune the predefined time for defining the recent period (i.e., $T$) to have the recent period of $\{16, 32, 64\}$ weeks, and set the first 5 epochs as the warm-up steps. 
In PIS task, we handle the class imbalance by controlling the scale of losses for the negative label, i.e., $y_{u,i}=0$, with a factor $\gamma$ within $\{1, 0.1, 0.01\}$.
We also use the recent half of the frequency bins $\textbf{b}_{u,i}$ to save the computation time as we observe that the past half of the frequency bins only have a modest effect to the recommendation accuracy.

\subsection{Recommendation Accuracy Comparison}
We tabulate the recommendation accuracy of the recommender systems in Table \ref{tab:perf}, and make the following observations. 1)
\MD{} significantly outperforms the baseline models including the general, user-centric, and item-centric sequential models on the 11 real-world datasets. This result indicates the effectiveness of incorporating the PIS, i.e., whether each user's interest in items will sustain beyond the training time, over various domains. 
We provide detailed analyses on \MD{} in the following sections to understand its behavior and ensure that the PIS is indeed crucial to enhance the recommendation accuracy.
2) The item-centric sequential model, i.e., CRIS, shows better performance than the user-centric models, e.g., LSAN, indicating the importance of the interest sustainability even though it is considered only at the item-level (i.e., non-personalized).
However, due to the characteristic of datasets, CRIS does not consistently outperform LSAN, e.g., on Digital Music and Yelp datasets. In contrast, \MD{} takes the benefits of both user- and item-centric models, resulting in a higher recommendation accuracy than the models in either category. 
3) General recommender systems, e.g., SML and SimpleX, reach or outperform the performance of the user- and item-centric sequential recommender systems without modeling the sequence signal from the users' consumption history. Despite their competitive performance, \MD{} consistently outperforms the general models, which implies the effectiveness of the PIS inferred from users' sequential consumption history.
In addition, our experiments show that the sequential models are not consistently superior compared to the recent general models even though it is known that the sequential models are more accurate than the general models.

\subsection{Model Behavior Comparison}
In this section, we analyze the behavior of the recommender systems to understand the benefit of \MD{} compared to the baseline models. Due to the space limitation, we report the average of HR@10 and nDCG@10 on the test data as the performance, but the results are similar when using either one. 
\subsubsection{\textbf{Elapsed Time Since Users' Last Consumption.}}
\begin{figure}[t]
	\centering
	\includegraphics[width=.99\linewidth]{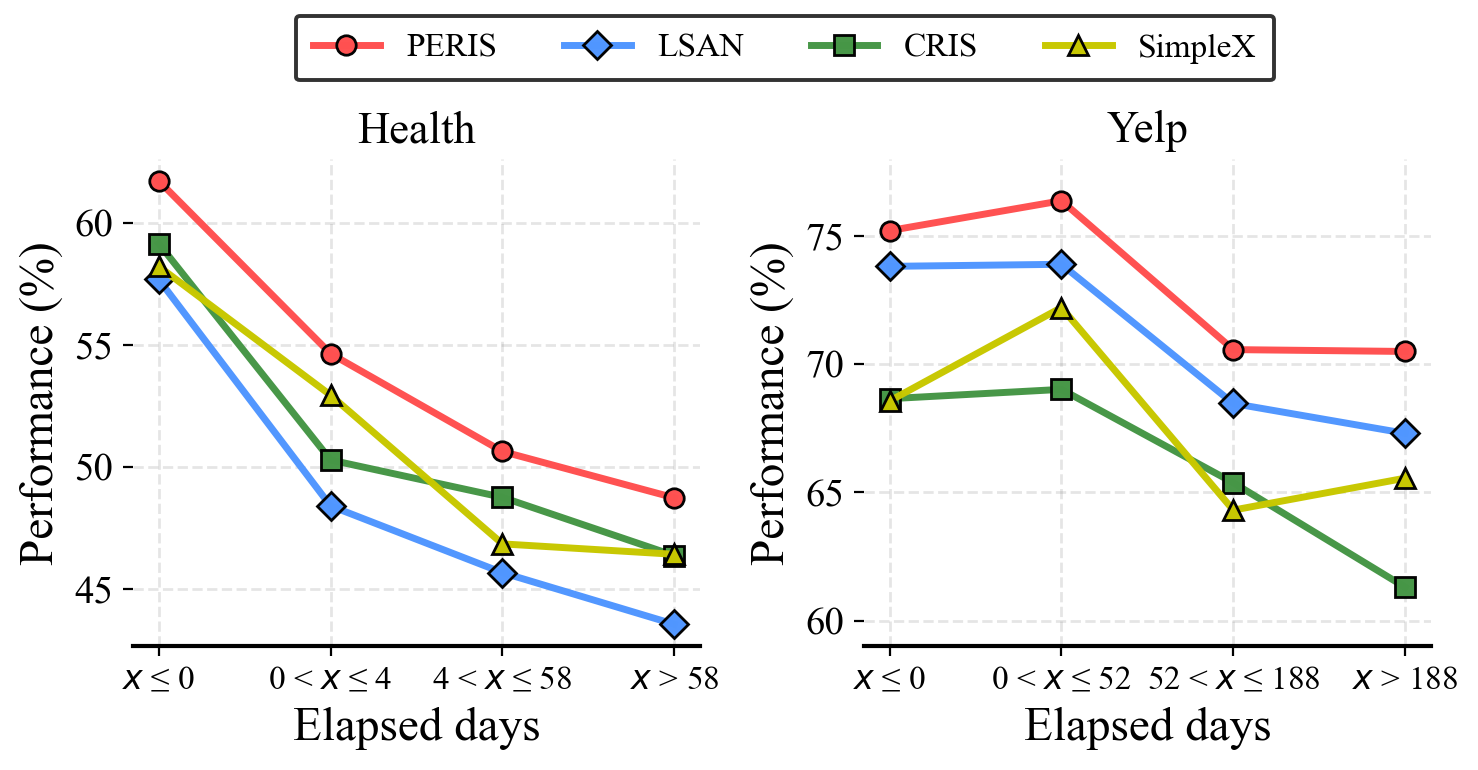}
	\caption{Performance over users' elapsed days since their last consumption. Performance: (HR@10+nDCG@10)/2.}
	\label{fig:elap}
\end{figure}

 In Figure \ref{fig:elap}, we report the recommendation accuracy of the models for different user groups divided by the elapsed time since their last consumption, e.g., $0 < x \le 4$ is the group of users who have their last consumption within 1 $\sim$ 4 days before the test time. To divide user groups, we compute the percentile of users' elapsed times since their last consumption and use 25-th, 50-th, and 75-th percentile as the threshold, e.g., 0, 4 and 58 on Health dataset, respectively. 
 
We observe that the performance of the models decreases when the elapsed time since their last consumption becomes long due to the prolonged absence of users' consumption. We have the following observations. 
1) Among the models, \MD{} consistently enhances the recommendation accuracy over different elapsed times compared to all the baseline models. 
2) The performance improvement of \MD{} over LSAN becomes larger as the elapsed time increases because LSAN depends on the \textit{next item prediction} that learns users' interest only up to their last consumption, while \MD{} learns users' recent interest by performing the \textit{PIS prediction}.
3) CRIS outperforms LSAN on the Health dataset as CRIS learns users' recent interest by predicting whether any user consumes each item in the recent period of the training time. 
However, CRIS is limited to learn users' recent \textit{personalized} interest as its prediction task is formulated as item-level. Thus, \MD{} surpasses CRIS thanks to the user-level prediction task, i.e., the PIS prediction. 
4) Compared to the sequential models, SimpleX shows inconsistent trends of performance over the elapsed time as it does not depend on the sequential information of users' consumption history.
However, \MD{} consistently outperforms SimpleX, signifying the importance of the sequential information in the form of the PIS. 
Thus, these observations imply that the PIS is beneficial to accurately infer users' interest compared to existing baseline models.

\subsubsection{\textbf{PIS in Other Recommender Systems}} 
\begin{figure}[t]
	\centering
	\includegraphics[width=.99\linewidth]{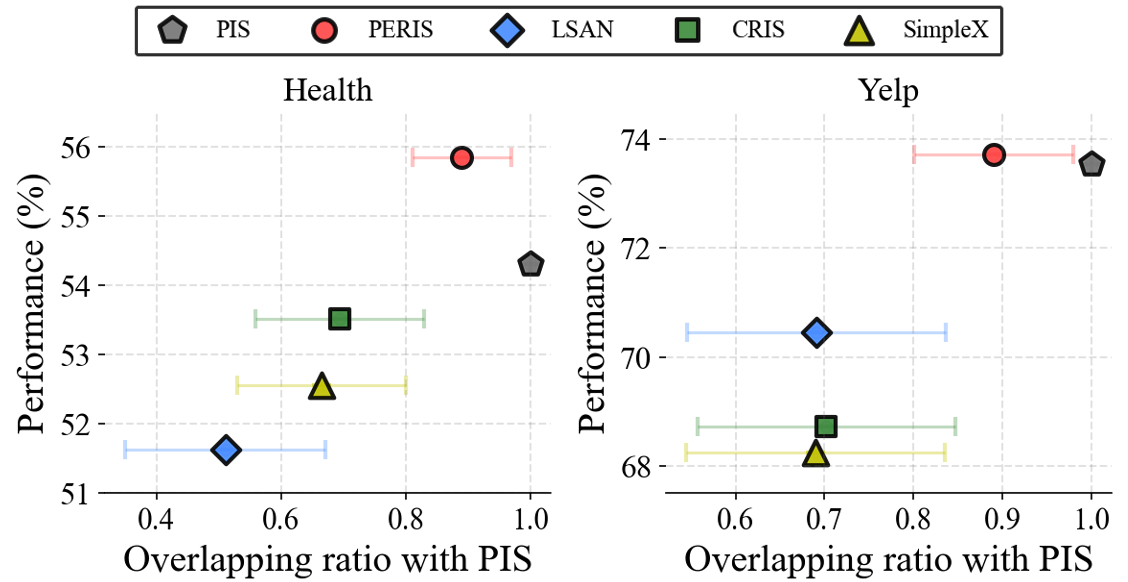}
	\caption{Performance over overlapping ratio between recommendation lists of recommender systems and PIS.  }
     \label{fig:overlap}
\end{figure}

We investigate whether other baseline models can capture the PIS or not (Figure \ref{fig:overlap}). We compute the overlapping ratio between the recommendation lists generated by other baseline models and the one generated based on the PIS score, i.e., $\mu \hat{y}_{u,i}^I + (1-\mu)\hat{y}_{u,i}^E$ from Equation \ref{eqn:score}, which is learned by \MD{}. In this experiment, we consider the top-10 items for each user in the test data as the recommendation list. From the result, we make several observations: First, the baseline models including user- and item-centric models show lower overlapping ratio with the recommendation list generated based on the PIS score than \MD{} does. Thus, their approaches to learn users' future interest cannot fully capture the PIS, while \MD{} captures the PIS thanks to the explicit modeling of the PIS. 
Second, depending solely on the PIS score can result in inaccurate recommendation, i.e., PIS on both datasets. This can be due to the label noise as the labels for the PIS task are not the ground-truth labels but the pseudo-labels. 
We argue that it is beneficial to address the label noise by incorporating users' preference score (i.e., $\hat{y}_{u,i}^P$ in Equation \ref{eqn:score}) from the classical preference learning along with the PIS scores.

\begin{figure}[t]
	\centering
	\includegraphics[width=.999\linewidth]{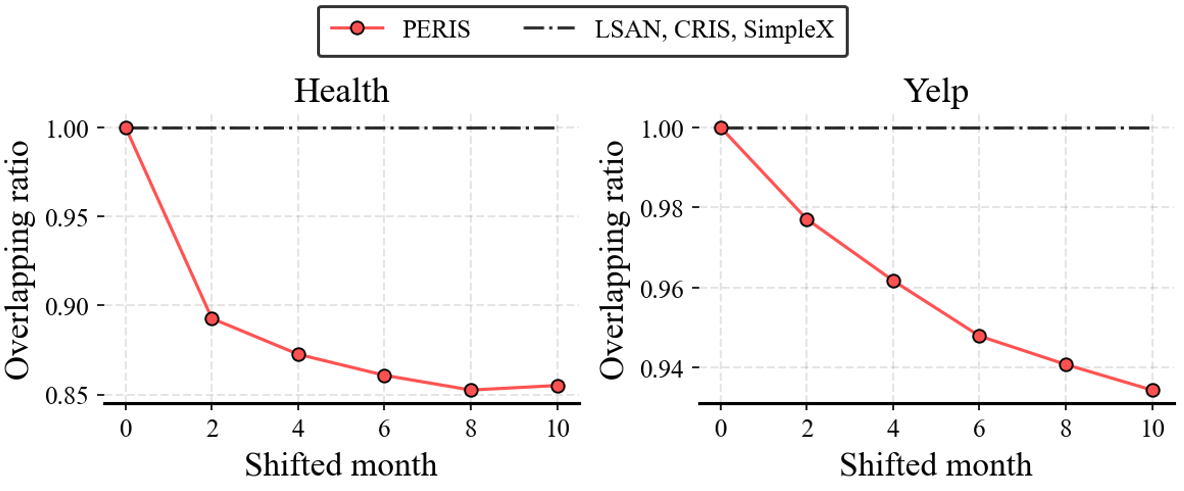}
	\caption{Overlapping ratio between original recommendation list and lists after shifting users' consumption history.}
	\label{fig:diverse}
\end{figure}

\subsubsection{\textbf{Sensitivity to Shifted Consumption History}}
To illuminate the benefit of \MD, we compare the sensitivity of the models to the temporal change in users' consumption history (Figure \ref{fig:diverse}). In this experiment, we shift users' consumption history, i.e., each user's whole sequence of consumed items, by month from its start time to the relative past or future without distinction between them, while maintaining the order and intervals of items. Then, we measure how the recommendations after shifting users' consumption history are changed from the original recommendations (i.e., shifted month $=0$) by computing the overlapping ratio between them. We consider top-10 items for each user  as the recommendation list. 
We observe that the general, user-centric, and item-centric baseline models (black dashed line) are invariant to the consumption history shift because they cannot differentiate the same consumption histories that start from different times. Specifically, LSAN only considers the order information in the consumption history, and CRIS does not utilize each user's consumption history.
In contrast, \MD{} differentiates consumption histories that start from different times thanks to the PIS with the feature $\textbf{b}_{u,i}$, which considers the time each user consumes items. Hence, \MD{} can produce more personalized recommendations than the state-of-the-art baseline models, which supports the superior performance of \MD{}.

\subsection{In-depth Model Analysis}
We provide experiments to understand the impact of each component of \MD{} with the ranking metrics on the validation data.

\begin{table}[t]
\centering
\small
\caption{Ablation study of \MD{}. Int. and Ext. denote intrinsic and extrinsic scheme.}
\resizebox{0.95\linewidth}{!}{
\begin{tabular}{ccc|cc|cc|cc}
\toprule
\multicolumn{3}{c}{Components} & \multicolumn{2}{c}{Tools} & \multicolumn{2}{c}{Toys} & \multicolumn{2}{c}{Yelp}        \\ \cmidrule(r){1-3} \cmidrule(lr){4-5}  \cmidrule(lr){6-7} \cmidrule(l){8-9}
\makecell{Int.}    & \makecell{Ext.} & \makecell{PIS}   & H@10   & N@10   & H@10   & N@10  & H@10   & N@10 \\ \midrule
\cmark        & \cmark  & \cmark           & \textbf{60.34}	&\textbf{39.16}	&\textbf{71.73}	&\textbf{51.36}	&\textbf{89.85}	&\textbf{59.53} \\
\xmark        & \cmark   & \cmark          & 59.39	&38.24	&69.55	&49.57	&87.79	&58.94 \\
\cmark       & \xmark  & \cmark            & 46.60	&25.03	&64.62	&44.12	&86.77	&53.96 \\ 
\xmark        & \xmark  & \cmark           & 46.41	&24.95	&64.25	&43.15	&\textcolor{black}{88.10}	&\textcolor{black}{54.10}  \\ 
\xmark        & \xmark  & \xmark           & 52.01	&31.96	&65.92	&46.23	&85.15	&53.87 \\ \bottomrule
\toprule
\multicolumn{3}{c|}{Last bin ($b_{u,i}^N$)}                            & 54.21	&33.76	&66.85	&47.97	&86.36	&54.28         \\ 
\multicolumn{3}{c|}{$-\, \textbf{e}_{u,i}$ (Eqn. \ref{eqn:pred})}        & \textbf{60.34}	&38.62	&71.40	&51.01	&86.77	&57.98         \\ 
\multicolumn{3}{c|}{$+\, \textbf{e}_{u,i} \,$ to Ext.}                      & 56.32	&33.31	&70.52	&49.10	&87.38	&54.50         \\ \bottomrule
\end{tabular}
}
\label{tab:abl_is}
\label{tab:abl_detail}
\end{table}

\subsubsection{\textbf{Ablation Study on Supplementation Schemes.}}
\label{sec:abl}
We first perform an ablation study to inspect the effect of the supplementation schemes of \MD{}.
From Table \ref{tab:abl_is}, we make the following observations: 1) the exclusion of either the intrinsic or extrinsic schemes (from second to fourth row in the table) degrades the recommendation accuracy due to the absence of the supplementation scheme for alleviating the sparsity of users' consumption history. Thus, both supplementation schemes are crucial to successfully capture the PIS for improving the recommendation accuracy, while the extrinsic scheme is more effective than the intrinsic scheme. 
2) The intrinsic scheme enhances the recommendation accuracy when it is used along with the extrinsic scheme (from first to second row in the table). This result suggests that both intrinsic and extrinsic should be incorporated together to achieve a higher recommendation accuracy. 
3) The model trained without the intrinsic and extrinsic schemes (fourth row in the table) generally produces a lower recommendation accuracy than the model trained only on the preference learning (fifth row in the table). The result reaffirms that the vanilla PIS prediction task suffers from the sparsity of users' consumption history, resulting in the failure of the PIS prediction task. Therefore, both supplementation schemes are vital to successfully perform the PIS prediction task. 

\subsubsection{\textbf{Ablation Study on Components.}}
We further provide the ablation study on the components of \MD{}. 
First, we study the effect of users' temporal consumption history $\textbf{b}_{u,i}$, which is the feature for the PIS prediction task. We use the most recent frequency bin $b_{u,i}^N$ instead of a sequence of the frequency bins, i.e., $\textbf{b}_{u,i}$. 
\MD{} that takes only the last bin $b_{u,i}^N$ (Last bin in Table \ref{tab:abl_detail}) shows the substantial degradation of recommendation accuracy.
Thus, we assert the necessity of users' sequential consumption pattern to predict the PIS. 
Second, we exclude the user-item joint representation for performing the PIS prediction task with the intrinsic scheme ($-\,\textbf{e}_{u,i}$ in Table \ref{tab:abl_detail}), and observe the degradation of recommendation accuracy in most cases. This result identifies the importance of the joint information of users and items to predict their interest in items beyond the training time with the intrinsic feature, i.e., $\hat{y}_{u,i}^I = f_s(\textbf{b}_{u,i}^I + \textbf{e}_{u,i})$. In contrast, the inclusion of the user-item representation $\textbf{e}_{u,i}$ to the extrinsic feature $\textbf{b}_{u,i}^E$ decreases the recommendation accuracy ($+\,\textbf{e}_{u,i}$ to Ext. in Table \ref{tab:abl_detail}), i.e., $\hat{y}_{u,i}^E = f_s(\textbf{b}_{u,i}^E + \textbf{e}_{u,i})$. We conjecture that the inclusion of the target user-item representation $\textbf{e}_{u,i}$ is noisy since the goal of the extrinsic scheme is to predict not a target user's interest but the other users' interest. These results suggest to include the joint representation $\textbf{e}_{u,i}$ only to the intrinsic scheme.

\begin{figure}
     \centering
     \begin{subfigure}[b]{0.475\linewidth}
         \centering
         \includegraphics[width=\textwidth]{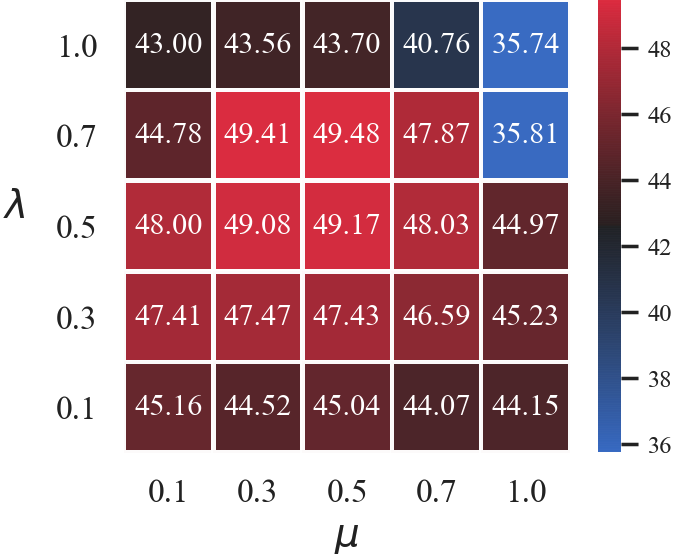}
         \caption{Tools}
         \label{fig:sense_train}
     \end{subfigure}
     \begin{subfigure}[b]{0.475\linewidth}
         \centering
         \includegraphics[width=\textwidth]{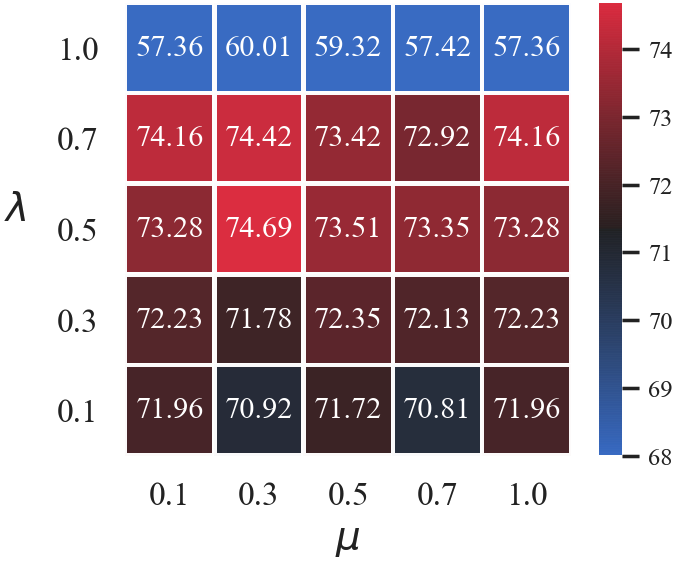}
         \caption{Yelp}
         \label{fig:sense_test}
     \end{subfigure}
     \caption{Sensitivity analysis on balancing parameters.}
     \label{fig:sensitivity}
\end{figure}

\subsubsection{\textbf{Sensitivity Analysis of Balancing Parameters.}}
In Figure \ref{fig:sensitivity}, we analyze the sensitivity of the balancing terms (i.e., $\lambda$ and $\mu$), which are used to balance among the losses in Equation \ref{eqn:final_loss}. We note that $\lambda$ balances between the losses for the PIS prediction task (i.e., $\mathcal{L}_I + \mathcal{L}_E$) and the preference learning (i.e., $\mathcal{L}_P$), and $\mu$ balances between the loss for the intrinsic scheme $\mathcal{L}_I $ and the loss for the extrinsic scheme $\mathcal{L}_E$. 
In this experiment, we use the average of HR@10 and nDCG@10 as the value for each combination. 
From the analysis (Figure \ref{fig:sensitivity}), we first observe that \MD{} trained only on the PIS task (i.e., $\lambda=1$) shows a substantial degradation of recommendation accuracy as the model learns users' interest based only on whether a user consumes items in the recent period or not. We speculate that, even though a user does not consume an item in the recent period, the user's interest in the item can sustain up to the future, e.g., a user has a long consumption period. Thus, to alleviate the label noise, it is better  to incorporate the preference learning (i.e., $\mathcal{L}_P$), which uses ground-truth labels (i.e., users' consumption), along with the PIS prediction task. The second observation is that the best value of $\mu$ is around 0.3. Thus, other liked-minded users' interest is essential to infer a target user's interest in items, which reconfirms the observation from the ablation study (\cref{sec:abl}).

\section{Conclusion}
In this work, we propose a recommender system that captures the personalized interest sustainability (PIS), indicating whether each user's interest in items will sustain beyond the training time, i.e., up to the test time. To obtain the PIS, we formulate the PIS prediction task, and devise the simple yet effective schemes to supplement users' sparse consumption history. 
Experiments show that the proposed model, i.e., \MD{}, enhances the recommendation accuracy compared to 10 baseline models on 11 real-world datasets. In addition, in-depth analysis reveals that \MD{} successfully captures the PIS while the baseline models do not capture the PIS, which is newly-introduced information.

\begin{acks}
This work was supported by IITP (No.2018-0-00584, SW starlab), (No.2019-0-01906, Artificial Intelligence Graduate School Program (POSTECH)), NRF grant (South Korea, No.2020R1A2B5B03097210, No.2021R1C1C1009081) funded by the Korea government (MSIT).
\end{acks}


\bibliographystyle{ACM-Reference-Format}
\balance
\bibliography{main}

\newpage

\end{document}